\documentclass[entropy,article,accept,moreauthors]{Definitions/mdpi} 

\graphicspath{{Definitions/}} 
\firstpage{1} 
\pubvolume{21}
\issuenum{4}
\articlenumber{380}
\pubyear{2019}
\copyrightyear{2019}
\history{Received: 24 December 2018; Accepted: 3 April 2019; Published: 8 April 2019}
\updates{yes} % If there is an update available, un-comment this line

\newcommand{\lam}{\lambda}
\newcommand{\R}{{\rm res}}
\newcommand{\AR}{{\rm ar}}
\newcommand{\re}{\mathop{\mathrm{Re}}}
\newcommand{\im}{\mathop{\mathrm{Im}}}

\Title{Time-Reversal Symmetry and Arrow of Time in Quantum Mechanics of Open Systems}

\Author{Naomichi Hatano~$^{1}$*\orcidA{} and Gonzalo Ordonez~$^{2}$\orcidB{}}

\AuthorNames{Naomichi Hatano and Gonzalo Ordonez}

\address{%
$^1$ \quad Institute of Industrial Science, the University of Tokyo,
5-1-5 Kashiwanoha, Kashiwa, Chiba 277-8574, Japan\\
$^2$ \quad Department of Physics and Astronomy, Butler University,
4600 Sunset Avenue, Indianapolis, IN 46208, USA;
gordonez@butler.edu
}

\corres{Correspondence: hatano@iis.u-tokyo.ac.jp; Tel.: +81-4-7136-6977}

\abstract{
It is one of the most important and long-standing issues of physics to derive the irreversibility out of a time-reversal symmetric equation of motion.
The present paper considers the breaking of the time-reversal symmetry in open quantum systems and the emergence of an arrow of time.
We claim that the time-reversal symmetric Schr\"{o}dinger equation can have eigenstates that break the time-reversal symmetry if the system is open in the sense that it has at least a countably infinite number of states.
Such eigenstates, namely the resonant and anti-resonant states, have complex eigenvalues.
We show that, although these states are often called ``unphysical'', they observe the probability conservation in a particular way.
We also comment that the seemingly Hermitian Hamiltonian is non-Hermitian in the functional space of the resonant and anti-resonant states, and hence there is no contradiction in the fact that it has complex eigenvalues.
We finally show how the existence of the states that break the time-reversal symmetry affects the quantum dynamics.
The dynamics that starts from a time-reversal symmetric initial state is dominated by the resonant states for $t>0$; this explains the phenomenon of the arrow of time, in which the decay excels the growth.
The time-reversal symmetry holds in that the dynamic ending at a time-reversal symmetric final state is dominated by the anti-resonant states for $t<0$.}

\begin{document}
%%%%%%%%%%%%%%%%%%%%%%%%%%%%%%%%%%%%%%%%%%

%%%%%%%%%%%%%%%%%%%%%%%%%%%%%%%%%%%%%%%%%%
%\setcounter{section}{-1} %% Remove this when starting to work on the template.

\section{Introduction}
All microscopic equations of motion, only with a slight exception of the weak interaction, have the time-reversal symmetry. 
It is therefore one of the biggest issues of physics to know how the irreversibility in thermodynamics and statistical physics, namely the ``arrow of time''~\cite{Eddington1948},  emerges out of the time-reversal-symmetric equations of motion. 

This issue was pointed out by Loschmidt to Boltzmann back in the 19th century~\cite{Bader01}, and has been addressed by many authors ever since.  It is safe to say that this problem is not fully resolved yet, but progress has been made. Some authors~\cite{Prigogine73, Prigogine81, Bohm89a, Petrosky97, Driebe1999} have proposed to extend the domain of classical or quantum operators to include complex eigenvalues, thus formulating a dynamics with broken-time symmetry.  Other authors have taken a more conventional approach and explained Loschmidt's paradox resorting to probabilistic arguments. For example, Evans and Searle formulated a fluctuation theorem where they quantified the probability of violations of the second law for classical systems~\cite{Evans02}. More recently, Maccone~\cite{Maccone09} proposed that processes that violate the second law are  unobservable; Vaccaro~\cite{Vaccaro16} suggested that the $T$ violation by some elementary particles is the ultimate root of the arrow of time, while Carroll~\cite{Carroll2010} suggested that the arrow of time comes from spontaneous eternal inflation, creating universes with low initial entropy.

Here, we review our recent work on this subject~\cite{Hatano14,Ordonez17a,Ordonez17b}, where we argued that an  arrow of time can emerge even in very simple quantum systems. We stress the fact that each \textit{solution} of time-reversal equation of motion can break the time-reversal symmetry if the system is open. %Is the italics necessary? 
%% Authors' Reply: I deleted italic style from two "can"s but otherwise, we would like to keep them italic.
Therefore, we can indeed have, in specific cases, the dynamics that seemingly breaks the time-reversal symmetry. This is a~well known fact to some researchers but unfortunately does not seem to be acknowledged widely. It~is our intention to emphasize and demonstrate the fact. 
We additionally claim that the initial-condition problem always chooses the states that break the time-reversal symmetry in a decaying way~\cite{Hatano14,Ordonez17a,Ordonez17b}.

In the present article, we restrict ourselves to quantum mechanics, but it should be easily generalized to other equations of motion. 
Section~\ref{sec2} shows that the time-reversal symmetry of the Schr\"{o}dinger equation does not exclude the existence of eigenstates that break the time-reversal symmetry. Such states must have complex eigenvalues, as also pointed out by other authors \cite{Sudarshan78,Bohm89a,Petrosky91,Petrosky97}.
We then demonstrate in Section~\ref{sec3} that open quantum systems indeed can have complex eigenvalues, and therefore can have eigenstates which break the time-reversal symmetry, namely the resonant and anti-resonant eigenstates.
However, why could a ``Hermitian'' Hamiltonian have a complex eigenvalue?
We stress in Section~\ref{sec4} that one needs to specify the functional space to prove the Hermiticity of a~Hamiltonian operator. 
Indeed, the Hamiltonian operator for the open quantum system is Hermitian in the standard Hilbert space with normalizable eigenfunctions, but \textit{not} Hermitian in the functional space that contains the states that break the time-reversal symmetry, and hence there is no contradiction if the symmetry-breaking eigenstates have complex eigenvalues~\cite{Hatano14,Ordonez17a,Ordonez17b,Petrosky91}.
Section~\ref{sec5} finally describes the essential difference between the breaking of the time-reversal symmetry and the existence of the arrow of time;
the proof of the latter needs one additional ingredient to the proof of the former.
The arrow of time means that we observe specific symmetry-breaking states more often than the other symmetry-breaking states, typically decaying states more often than growing states.
We show that the resonant states indeed excel the anti-resonant states in the initial-condition problem~\cite{Hatano14,Ordonez17a,Ordonez17b}.
The time-reversal symmetry of the Schr\"{o}dinger equation only guarantees that the anti-resonant states excel in the final-condition problem.
Section~\ref{sec6} is devoted to a summary.

\section{Time-Reversal Operator}
\label{sec2}

The Schr\"{o}dinger equation 
\begin{align}\label{eq10}
i\hbar \frac{d}{dt}|\Psi\rangle=H|\Psi\rangle,
\end{align}
the most used quantum-mechanical equation of motion, is time-reversal symmetric if the Hamiltonian $H$ commutes with the time-reversal operator $T$; in quantum mechanics, the time-reversal operator consists of two combined operations of negating the time $t$ and the complex conjugation:
\begin{align}
t\to -t; \quad
i\to-i.
\end{align}
Applying the time-reversal operator $T$ from the left of the Schr\"{o}dinger Equation~\eqref{eq10} with the use of the commutability
\begin{align}\label{eq20}
[H,T]=0,
\end{align}
we have
\begin{align}
i\hbar \frac{d}{dt}\left(T|\Psi\rangle\right)=H\left(T|\Psi\rangle\right),
\end{align}
which means that the time-reversed state $T|\Psi\rangle$ indeed obeys the same equation as the original state $|\Psi\rangle$ does.
One might therefore expect that if we watched a video of the dynamics of a quantum-mechanical state, we would not be able to tell whether the video was rolling forward or backward.

An important fact to note here is that the time-reversal operator $T$ is an anti-linear operator, \textit{not} a~linear operator.
Because of this fact, the two operators $H$ and $T$, which commute with each other, may not share common eigenstates.
Let us demonstrate the fact in an elementary way.
Suppose that we know an eigenstate and its eigenvalue of the Hamiltonian $H$:
\begin{align}\label{eq50}
H|\psi\rangle = E|\psi\rangle.
\end{align}
Applying the time-reversal operation from the left, we have
\begin{align}\label{eq40}
TH|\psi\rangle=TE|\psi\rangle.
\end{align}
The left-hand side is equal to $HT|\psi\rangle$ because of the commutability~\eqref{eq20}.
The right-hand side \textit{would} be equal to $ET|\psi\rangle$ if $T$ were a linear operator;
then we would conclude that the state $T|\psi\rangle$ is parallel to the state $|\psi\rangle$ provided that its eigenvalue $E$ does not have degeneracy. 
This would mean that the solution $|\psi\rangle$ is time-reversal symmetric, that is, it is an eigenstate of the time-reversal operator $T$.
This is the fact that we all learn in the class of elementary quantum mechanics; 
if there are two \textit{linear} operators $A$ and $B$ that commute with each other as in $[A,B]=0$, they share eigenstates.
It is, however, not stressed enough that this statement generally applies to linear operators only.

In fact, the time-reversal operator $T$ is an anti-linear operator. Therefore, the right-hand side of Equation~\eqref{eq40} is equal to $E^\ast T|\psi\rangle$, and hence Equation~\eqref{eq40} is followed by
\begin{align}\label{eq70}
H\left(T|\psi\rangle\right)=E^\ast\left(T|\psi\rangle\right).
\end{align}

This means either of the following two cases:
\begin{enumerate}[leftmargin=*,labelsep=4.9mm]
\item The eigenvalue $E$ is real as in $E^\ast=E$. The state $T|\psi\rangle$ satisfies the same equation as  the original state $|\psi\rangle$ does with the same eigenvalue as is seen in Equations~\eqref{eq50} and~\eqref{eq70}, and hence they are proportional to each other provided that its eigenvalue $E$ does not have degeneracy. The two operators $H$ and $T$ share the eigenstate $|\psi\rangle$, which is therefore time-reversal symmetric.
\item The eigenvalue $E$ is complex as in $E^\ast\neq E$. Equations~\eqref{eq50} and~\eqref{eq70} are different in their eigenvalues; they now mean that if there is an eigenstate $|\psi\rangle$ with a complex eigenvalue $E$, there must be a~time-reversed eigenstate $T|\psi\rangle$ with a complex-conjugate eigenvalue $E^\ast$.
\end{enumerate}

We thereby conclude here that if the Hamiltonian has a complex eigenvalue, the corresponding eigenstate breaks the time-reversal symmetry. 
The time-reversal symmetry of the original equation of motion reflects in the fact that the symmetry-breaking eigenstates can appear always in a complex-\ conjugate pair.

The next issue is in what case the Hamiltonian does have a complex eigenvalue. We will show in the next section that the open quantum system is the case.

%The introduction should briefly place the study in a broad context and highlight why it is important. It should define the purpose of the work and its significance. The current state of the research field should be reviewed carefully and key publications cited. Please highlight controversial and diverging hypotheses when necessary. Finally, briefly mention the main aim of the work and highlight the principal conclusions. As far as possible, please keep the introduction comprehensible to scientists outside your particular field of research. Citing a journal paper \cite{ref-journal}. And now citing a book reference \cite{ref-book}. Please use the command \citep{ref-journal} for the following MDPI journals, which use author-date citation: Administrative Sciences, Arts, Econometrics, Economies, Genealogy, Humanities, IJFS, JRFM, Languages, Laws, Religions, Risks, Social Sciences.
 
%%%%%%%%%%%%%%%%%%%%%%%%%%%%%%%%%%%%%%%%%%
\section{Complex Eigenvalues}
\label{sec3}

We now demonstrate that an \textit{open} quantum system can have complex eigenvalues even if the Hamiltonian is (seemingly) Hermitian~\cite{Hatano14,Ordonez17a,Ordonez17b}.
The simplest model that does have complex eigenvalues may be a one-dimensional tight-binding model, which is often analyzed in condensed-matter physics.
It is also a tutorial to demonstrate that the openness with a countably infinite number of states is sufficient to produce complex eigenvalues. 

\begin{figure}[H]
\centering
\includegraphics[width=0.8\textwidth]{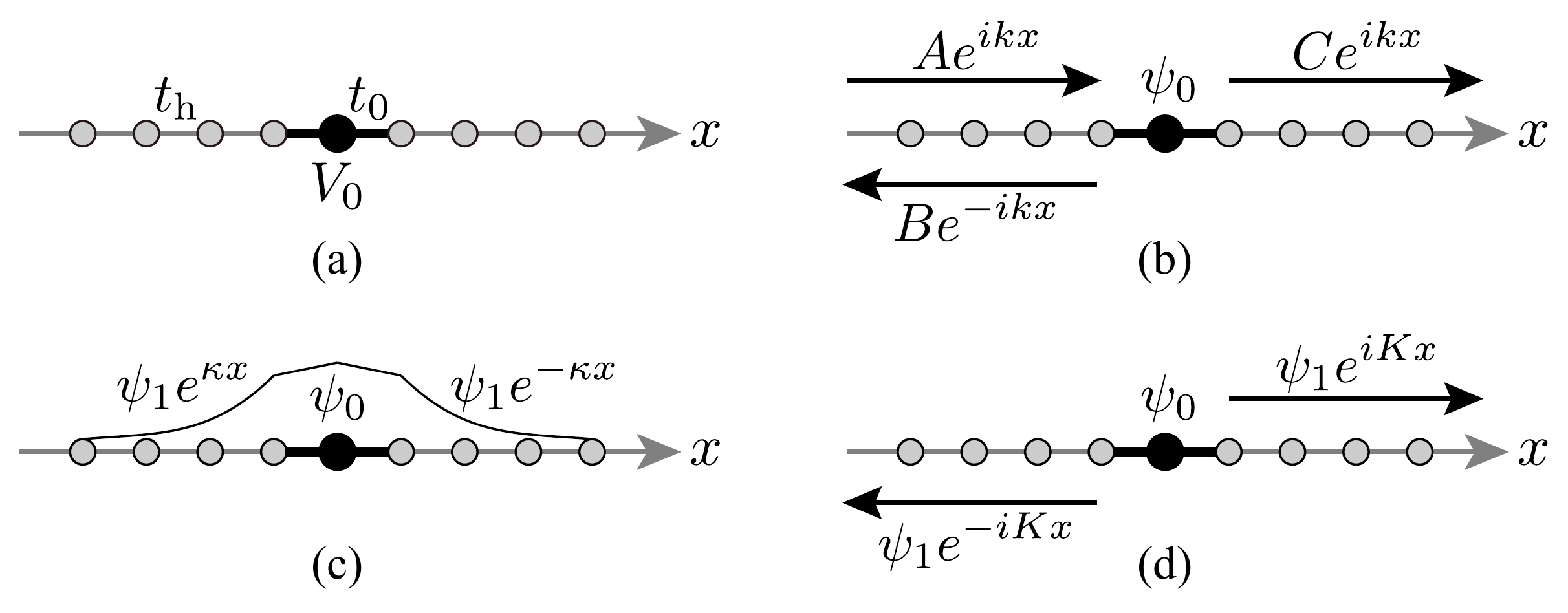}
\caption{({\bf a}) An infinite chain of atoms with an impurity at $x=0$.
({\bf b}) A scattering eigenstate. ({\bf c})~A~bound state. ({\bf d}) A general discrete eigenstate under the Siegert boundary condition.}
\label{fig1}
\end{figure}

Consider an infinite chain of atoms, on which an electron hops from one atom at $x$ to the neighboring ones $x\pm1$ with $x$ being an integer.
[Note that we put here the lattice constant (the distance between the neighboring sites) $a$ to unity;
all quantities of the order of distance will be implicitly given in the unit of $a$ hereafter.{]}
We put one impurity atom at $x=0$, which has a real-valued chemical potential of $V_0$; see Figure~\ref{fig1}a~\cite{Ordonez17b}.
We ignore the electron-electron interaction. 

The quantum-mechanical Hamiltonian that expresses the situation is
\begin{align}\label{eq80}
H=&-t_\mathrm{h}\left(\sum_{x=-\infty}^{-2}+\sum_{x=+1}^{+\infty}\right)
\Bigl(|x+1\rangle \langle x|+|x\rangle \langle x+1|\Bigr)
\nonumber\\
&-t_0\Bigl(|0\rangle\langle -1|+|-1\rangle\langle 0|\Bigr)-t_0\Bigl(|1\rangle\langle0|+|0\rangle\langle1|\Bigr)
\nonumber\\
&+V_0|0\rangle \langle 0|,
\end{align}
where $|x\rangle$ denotes the state in which the electron resides on the atom at $x$, while $t_\mathrm{h}$ and  $t_0$ are real-valued constants.
We here fix the hopping element $t_\mathrm{h}=-\langle x+1|H|x\rangle=-\langle x|H|x+1\rangle$ for $x\neq-1,0$ to unity for simplicity; all quantities of the order of energy will be implicitly in unit of $t_\mathrm{h}$ hereafter. 
We are now left with the two parameters $t_1= -\langle\pm1|H|0\rangle= -\langle0|H|\pm1\rangle \in\mathbb{R}$ and $V_0=\langle 0|H|0\rangle\in\mathbb{R}$.
The~Hamiltonian is seemingly Hermitian;
we will nonetheless demonstrate below that the Hamiltonian can have complex eigenvalues under a specific boundary condition called the Siegert boundary condition~\cite{Siegert39,Peierls59,Landau77,Hatano08}.

The Hamiltonian~\eqref{eq80} has the eigenvectors of both a continuous spectrum and point spectra, the latter of which contain the ones with complex eigenvalues.
Meanwhile, the former contains the usual scattering states of the form shown in Figure~\ref{fig1}b:
\begin{align}\label{eq90}
|\psi_\mathrm{s}\rangle
=\sum_{x=-\infty}^{-1}\Bigl(Ae^{ikx}+Be^{-ikx}\Bigr)|x\rangle
+\psi_0|0\rangle
+\sum_{x=+1}^{+\infty}Ce^{ikx}|x\rangle.
\end{align}
with the continuous eigenvalue varying in the energy range $-2\leq E\leq 2$; 
in condensed-matter physics this is called the energy band, which is described as
%\begin{align}
$E=-2\cos k$
%\end{align}
in terms of the wave number $k$ varying in the range $-\pi<k\leq \pi$.
This range of the wave number is called the first Brillouin zone; it is the restriction that comes from the discreteness of the lattice.
The latter class of the eigenvectors of point spectra include the usual bound state around the impurity atom at $x=0$ of the form shown in Figure~\ref{fig1}c:
\begin{align}\label{eq110}
|\psi_\mathrm{b}\rangle
=\sum_{x=-\infty}^{-1}\psi_1e^{\kappa x}|x\rangle
+\psi_0|0\rangle
+\sum_{x=+1}^{+\infty}\psi_1e^{-\kappa x}|x\rangle
\end{align}
with a discrete eigenvalue (a point spectrum)
%\begin{align}
{$E=-2\cosh \kappa$},
%\end{align}
where we can determine a positive number {$\kappa$} in terms of the parameters $t_0$ and $V_0$ by solving the time-independent Schr\"{o}dinger Equation~\eqref{eq50}.  

The general boundary condition for the eigenvectors of point spectra, which is called the Siegert boundary condition~\cite{Siegert39,Peierls59,Landau77,Hatano08}, is shown in Figure~\ref{fig1}d:
\begin{align}\label{eq130}
|\psi_\mathrm{p}\rangle
=\sum_{x=-\infty}^{-1}\psi_1e^{-i K x}|x\rangle
+\psi_0|0\rangle
+\sum_{x=+1}^{+\infty}\psi_1e^{i K x}|x\rangle,
\end{align}
where the wave number $K$ is generally complex with its real part restricted to the range $-\pi<\re K \leq \pi$.
Notice that the bound state~\eqref{eq110} belongs to the class of the point-spectral solutions~\eqref{eq130}, having a pure imaginary wave number {$K=i\kappa$} with {$\kappa>0$}.
The Siegert boundary condition~\eqref{eq130} meanwhile represents only outgoing waves when $\re K>0$ and only incoming waves when $\re K<0$.

To find the general eigenvector with a point spectrum, we insert Equation~\eqref{eq130} into the time-independent Schr\"{o}dinger Equation~\eqref{eq50}~\cite{Ordonez17b}.
%The left-hand side is given by
%\begin{align}
%H|\psi_\mathrm{b}\rangle
%=&\sum_{x=-\infty}^{-2}\Bigl(e^{\kappa x}|x+1\rangle+|x-1\rangle\Bigr)
%+\sum_{x=+1}^{+\infty}e^{-\kappa x}\Bigl(|x+1\rangle+|x-1\rangle\Bigr)
%\nonumber\\
%&-t_0\left(e^{\kappa}|0\rangle+\psi_0|-1\rangle\right)
%+V_0|0\rangle
%\end{align}
%to be equated to $E|\psi_\mathrm{b}\rangle$.
For $y\leq -2$ and $y\geq 2$, we respectively have
\begin{align}
\langle y|H|\psi_\mathrm{p}\rangle
&=-\psi_1\Bigl[e^{-iK (y-1)}+e^{-iK(y+1)}\Bigr]
=-2\psi_1e^{-iK y}\cos K,
\\
\langle y|H|\psi_\mathrm{p}\rangle
&=-\psi_1\Bigl[e^{iK (y-1)}+e^{iK(y+1)}\Bigr]
=-2\psi_1e^{iK y}\cos K,
\end{align}
both of which is to be equated to
\begin{align}
E\langle y|\psi_\mathrm{p}\rangle=E\psi_1e^{iK|y|}
\end{align}
for $|y|\geq2$. We thus find the eigenvalue to be
\begin{align}
E=-2\cos K.
\end{align}
We can relate $K$ to $t_0$ and $V_0$ by computing the matrix elements
\begin{align}
\langle 0| H |\psi_\mathrm{p}\rangle
&=-2t_0e^{iK}\psi_1+V_0\psi_0,
\\
\langle 1| H |\psi_\mathrm{p}\rangle
&=-t_0\psi_0-e^{2iK}\psi_1,
\end{align}
which we equate to
\begin{align}
E\langle 0|\psi_\mathrm{p}\rangle&=-2\psi_0\cos K,
\\
E\langle 1|\psi_\mathrm{p}\rangle&=-2\psi_1e^{iK}\cos K,
\end{align}
respectively. We thus have the matrix equation
\begin{align}\label{eq220}
\begin{pmatrix}
V_0 & -2t_0 \\
-t_0 & -e^{iK} 
\end{pmatrix}
\begin{pmatrix}
\psi_0\\
e^{iK}\psi_1
\end{pmatrix}
=-2\cos K
\begin{pmatrix}
\psi_0\\
e^{iK}\psi_1
\end{pmatrix},
\end{align}
which is followed by the secular equation
\begin{align}\label{eq230}
\det\begin{pmatrix}
V_0+2\cos K & -2t_0 \\
-t_0 & e^{-iK}
\end{pmatrix}
=0.
\end{align}
Note that we had two unknown variables $K$ and $\psi_1/\psi_0$, for which we have two equations in Equation~\eqref{eq220}. We thus know that we have always discrete solutions. 

It is convenient to use
\begin{align}
\lambda=e^{iK}
\end{align}
to solve the secular Equation~\eqref{eq230}. We thereby have
%\begin{align}
%\left(V_0+\lambda+\frac{1}{\lambda}\right)\frac{1}{\lambda}-2{t_1}^2=0,
%\end{align}
%which is followed by 
a quadratic equation with respect to $\lambda$ of the form
\begin{align}
(1-2{t_0}^2)\lambda^2+V_0\lambda +1=0,
\end{align}
whose solutions are
\begin{align}
\label{eq:lambdap}
\lambda_\mathrm{p}=\frac{-V_0\pm\sqrt{{V_0}^2-4(1-2{t_0}^2)}}%
{2(1-2{t_0}^2)}.
\end{align}
We thus arrive at two discrete solutions with the eigen-wave-number and the eigen-energy given in the forms
\begin{align}
K_\mathrm{p}=-i\log \lambda_\mathrm{p},
\quad
E_\mathrm{p}=-2\left(\lambda_\mathrm{p}+\frac{1}{\lambda_\mathrm{p}}\right).
\end{align}

\begin{figure}[H]
\centering
\includegraphics[width=0.35\textwidth]{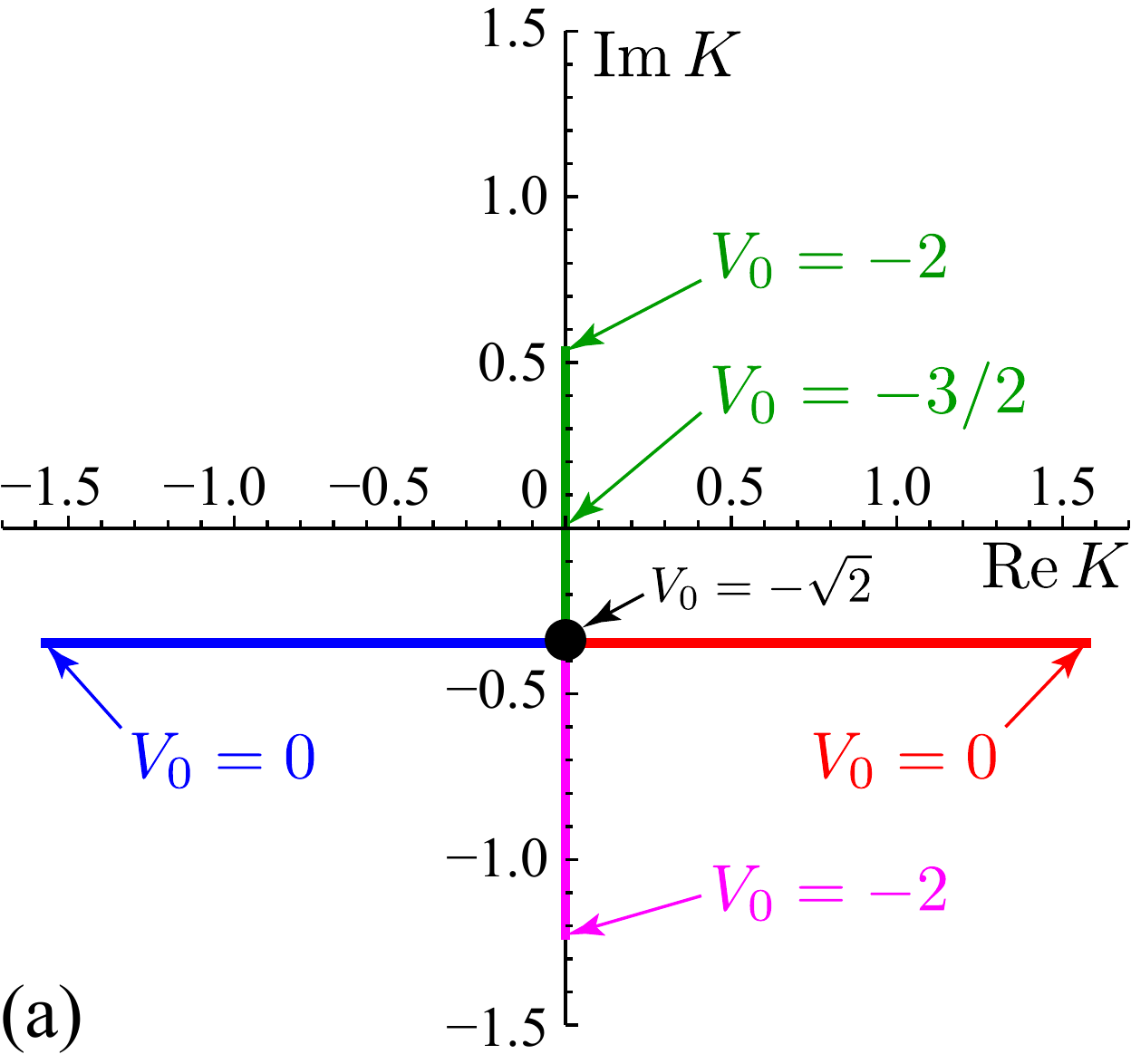}
\hspace{0.05\textwidth}
\includegraphics[width=0.35\textwidth]{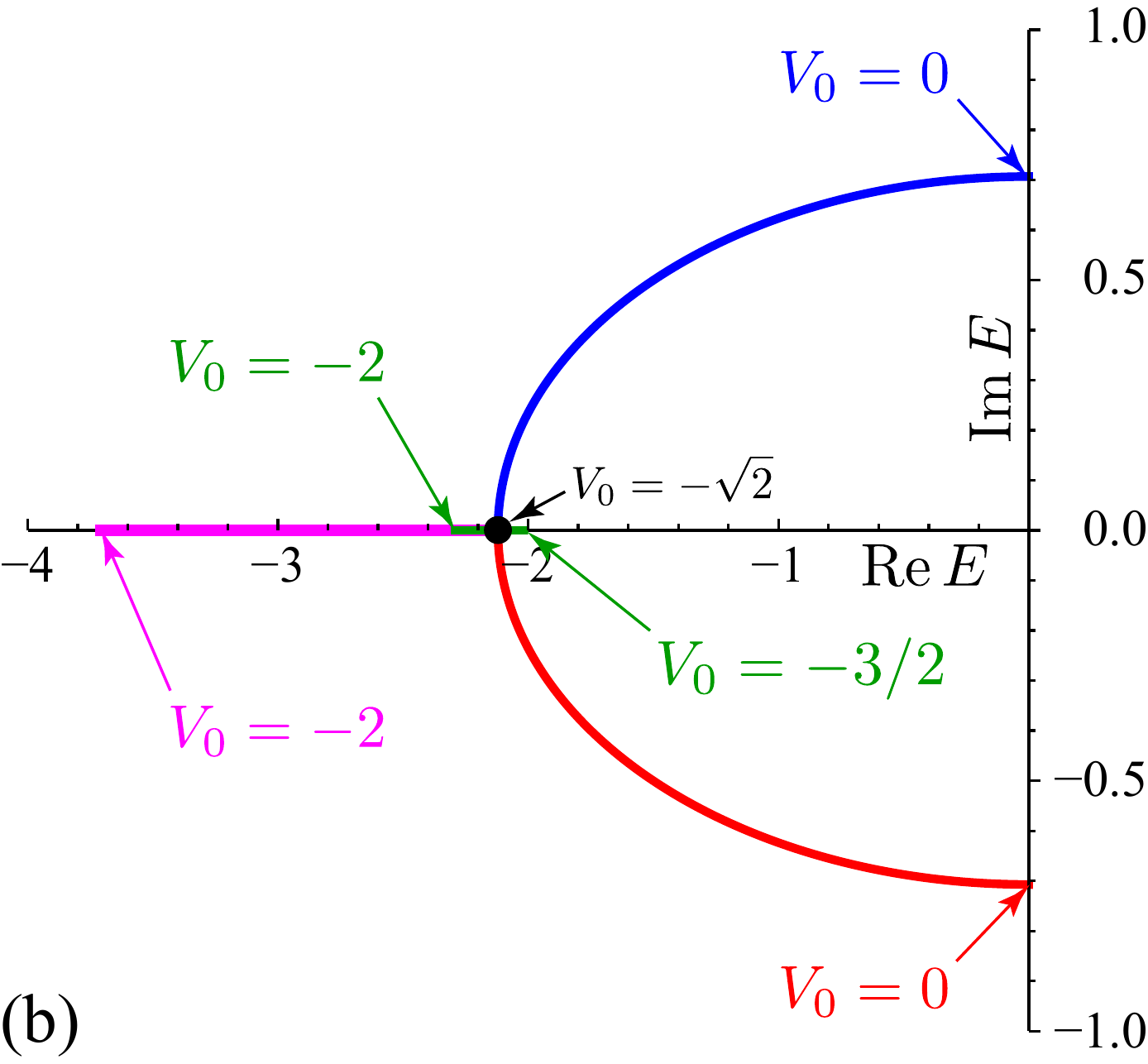}
\vspace{0.5\baselineskip}
\caption{The parametric plot of ({\bf a}) $K_\mathrm{p}$ and ({\bf b}) $E_\mathrm{p}$.
We fixed $t_0$ to $1/2$ and varied $V_0$ from $-2$ to $0$.
In~addition to these point spectra, the continuous spectrum $E=-2\cos k$ of the scattering states exists over the ranges of $-\pi\leq K\leq \pi$ and $-2\leq E\leq 2$ in (\textbf{a}) and (\textbf{b}), respectively.
In~(\textbf{a}), starting from $V_0=-2$, the bound state (green line) turns into an anti-bound state at $V_0=-3/2$ and coalesces with another anti-bound state (magenta line) at the exceptional point at $V_0=-\sqrt{2}$, branching out to a~pair of a resonant state (red line) and an anti-resonant state (blue line).
In~(\textbf{b}), the bound state (green line) moves up to the band edge $E=-2$ at $V_0=-3/2$ and doubles back to the exceptional point at $V_0=-\sqrt{2}$ to coalesce with the other anti-bound state (magenta line), branching out to the pair of the resonant (red line) and anti-resonant (blue line) states.}
\label{fig2}
\end{figure}

See Figure~\ref{fig2} for the parameter dependence of $K_\mathrm{p}$ and $E_\mathrm{p}$.
In the complex eigen-wave-number plane of $K$ in Figure~\ref{fig2}a:
the state on the positive imaginary axis is nothing but the bound state~\eqref{eq110} because {$K_\mathrm{p}=i\kappa$} with {$\kappa>0$};
the states on the negative imaginary axis is called anti-bound states, which we do not elaborate more here;
the state in the fourth quadrant is a resonant state, while the one in the third quadrant is an anti-resonant state.
The resonant state has only outgoing waves because $\re K>0$ in Equation~\eqref{eq130}, whereas the anti-resonant states has only incoming waves because $\re K<0$.
Both of these states have negative imaginary parts of $K_\mathrm{p}$, which we will focus on in the next section.

In the complex eigen-energy plane of $E$ in Figure~\ref{fig2}b,
the bound and anti-bound states have real eigenvalues, while
the resonant and anti-resonant states have complex eigenvalues, the former in the lower half ($\im E_\mathrm{p}<0$) and the latter in the upper half ($\im E_\mathrm{p}>0$).
They are in a complex-conjugate pair;
one of the pair is the time-reversed state of the other, as described in item~2 below Equation~\eqref{eq70}.

To summarize the above, open quantum systems indeed have complex eigenvalues, which correspond to eigenstates that break the time-reversal symmetry.
Note that the resonant state, the state~\eqref{eq130} with a positive real part of the wave number $K$, can exist only when the system is open;
in~a closed system, the out-going waves would bounce back at the boundaries, which would mix the resonant state with the anti-resonant state and reduce them to a standing wave.

A natural question is in order, however.
According to what we all learn, because the Hamiltonian~\eqref{eq80}  is Hermitian, it should have only real eigenvalues, should it not?
We will answer this question in the next section by stressing that the Hamiltonian of the open quantum system is Hermitian in the functional space of the normalizable eigenstates, but it is \textit{not} in the functional space that contains the resonant and anti-resonant eigenstates, and therefore it is indeed legitimate for these eigenstates to have complex eigenvalues.

\section{Why Complex Eigenvalues for ``Hermitian'' Hamiltonian?}
\label{sec4}

Let us first point out that the states in the upper half of the complex eigen-wave-number plane $K$, namely the bound states~\eqref{eq110}, are normalizable and the states on its real axis, namely the scattering states~\eqref{eq90}, are box-normalizable, whereas the states in the lower half of the complex eigen-wave-number plane $K$, namely the resonant and anti-resonant states~\eqref{eq130} (as well as the anti-bound states) are, in~the usual sense, unnormalizable.
Since the imaginary part of the eigen-wave-number $K_\mathrm{p}$ of the resonant and anti-resonant states is negative, the eigenstate~\eqref{eq130} spatially diverges: $e^{iK|x|}\to\infty$ as $x\to\pm\infty$.
These states are thereby often called ``unphysical states'' being not highly appreciated.
We will nonetheless stress in the present section that the unnormalizability is actually essential in the conservation of the ``norm'' that we define in a partial space.

Before that, let us first show that the Hamiltonian~\eqref{eq80} is \textit{not} Hermitian for the resonant and anti-resonant states~\cite{Hatano08,Hatano09,Hatano10}.
The Hermiticity of the Hamiltonian operator $H$ is often defined as follows:
the operator $H$ is Hermitian 
if and only if the matrix element $\langle \varphi|H|\varphi\rangle$ is real for an ``arbitrary'' state $|\varphi\rangle$.
%% Authors' Reply: Here we changed italic to the quotation marks.
The more precise statement, however, is the following:
\begin{Proposition}
If and only if the matrix element $\langle \varphi|H|\varphi\rangle$ is real, the Hamiltonian operator $H$ is Hermitian in the specific functional space from which $|\varphi\rangle$ is chosen arbitrarily.
\end{Proposition}

In order to check the Hermiticity of the Hamiltonian~\eqref{eq110}, let us calculate the matrix element $\langle \varphi|H|\varphi\rangle$ for an arbitrary state $|\varphi\rangle$.
Let us assume the expansion of the state as in
\begin{align}\label{eq260}
|\varphi\rangle=\sum_{x=-\infty}^{+\infty}\varphi_x|x\rangle.
\end{align}
The matrix element is then given by
\begin{align}\label{eq270}
\langle \varphi |H|\varphi\rangle
=&
-\sum_{x=-\infty}^{-2}{\varphi_x}^\ast \left(\varphi_{x-1}+\varphi_{x+1}\right)
\nonumber\\
&-{\varphi_{-1}}^\ast \left(\varphi_{-2}+t_0\varphi_0\right)
-t_0{\varphi_{0}}^\ast \left(\varphi_{-1}+\varphi_{1}\right)+V_0|\varphi_0|^2
-{\varphi_{1}}^\ast \left(t_0\varphi_{0}+\varphi_{2}\right)
\nonumber\\
&-\sum_{x=2}^{+\infty}{\varphi_x}^\ast \left(\varphi_{x-1}+\varphi_{x+1}\right).
\end{align}
One might transform this by means of resummation to 
\begin{align}\label{eq280}
\langle \varphi |H|\varphi\rangle
=&-\sum_{x=-\infty}^{-2}
\left({\varphi_x}^\ast\varphi_{x+1}+{\varphi_x}{\varphi_{x+1}}^\ast\right)
\nonumber\\
&-t_0\left({\varphi_{-1}}^\ast\varphi_0+\varphi_{-1}{\varphi_0}^\ast\right)
-t_0\left({\varphi_{0}}^\ast\varphi_1+\varphi_{0}{\varphi_1}^\ast\right)
+V_0|\varphi_0|^2
\nonumber\\
&-\sum_{x=+1}^{+\infty}
\left({\varphi_x}^\ast\varphi_{x+1}+{\varphi_x}{\varphi_{x+1}}^\ast\right),
\end{align}
and thereby conclude that it would be real when the parameters $t_0$ and $V_0$ are real.
The resummation is indeed legitimate if $|\varphi_{x}|$ vanishes as $|x|\to\infty$ quickly enough as in the bound state, and hence the expansion~\eqref{eq260} is convergent.
It is also legitimate if $|\varphi_x|$ is a plane wave $e^{ikx}$ with $k\in\mathbb{R}$ as $|x|\to\infty$ as in the scattering states.
It is \textit{not}, however, if $|\varphi_x|$ diverges as $|x|\to\infty$ as in the resonant and anti-resonant states, and the expansion~\eqref{eq260} is divergent and unnormalizable.
Indeed, for the states of the form~\eqref{eq130}, the summands in Equations~\eqref{eq270} and~\eqref{eq280} for $x\geq 1$ are respectively given by
using $\varphi_x=\psi_1e^{iK}$ for $x\geq 1$ as in
\begin{align}\label{eq290}
{\varphi_x}^\ast\left(\varphi_{x-1}+\varphi_{x+1}\right)
&=e^{-iK^\ast x+iK(x-1)}+e^{-iK^\ast x+iK(x+1)}
=\left(e^{iK}+e^{-iK}\right)e^{- 2x\im K}
\\
{\varphi_x}^\ast \varphi_{x+1}+\varphi_x{\varphi_{x+1}}^\ast&=e^{-iK^\ast x+iK(x+1)}+e^{-iK^\ast( x+1)+iKx}
=\left(e^{iK}+e^{-iK^\ast}\right)e^{- 2x \im K}
\end{align}
The difference of the two would decay out for $\im K>0$ as $x\to\infty$ if $K$ were complex.
The difference grows, however, for $\im K<0$ and hence the resummation is not legitimate~\cite{Hatano08,Hatano09,Hatano10}.
The former~\eqref{eq290} is indeed complex for a complex value of $K$.

%the state~\eqref{eq130}. A somewhat tedious algebra produces
%\begin{align}
%\langle\psi|H|\psi\rangle
%=&\sum_{x=-\infty}^{-2}\left(-2|\psi_1|^2e^{iK^\ast x-iKx} \cos K\right)
%+{\psi_1}^\ast e^{-iK^\ast}\left(-t_0\psi_0-e^{2iK}\psi_1\right)
%+{\psi_0}^\ast \left(-2t_0e^{iK}\psi_1+V_0\psi_0\right)
%\nonumber\\
%&+{\psi_1}^\ast e^{-iK^\ast}\left(-t_0\psi_0-e^{2iK}\psi_1\right)
%+\sum_{x=+2}^{+\infty}\left(-2|\psi_1|^2e^{-iK^\ast x+iKx} \cos K\right)
%\nonumber\\
%=&-4|\psi_1|^2\cos K \sum_{x=+2}^{+\infty}e^{-2\mathop{\mathrm{Im}K}|x|}
%-2t_0\psi_0{\psi_1}^\ast e^{-iK^\ast}-2 |\psi_1|^2e^{2iK-iK^\ast}
%-2t_0{\psi_0}^\ast\psi_1e^{iK}+V_0|\psi_0|^2
%\nonumber\\
%=&-2\cos K\left(|\psi_0|^2+ |\psi_1|^2 \sum_{{x=-\infty}\atop{x\neq0}}^{+\infty}e^{-2\mathop{\mathrm{Im}K}|x|}\right)
%=-2\cos K \langle\psi|\psi\rangle,
%\end{align}
%in the last line of which we took full advantage of Equation~\eqref{eq220}.
%This element is real when $|\psi\rangle$ is a bound state, for which $K$ is real and $\langle\psi|\psi\rangle$ is finite.
%It is, however, a complex number times infinity when $|psi\rangle$ is a resonant or anti-resonant state, for which $K$ is complex and the norm diverges.

This answers the question made at the end of the previous section: ``because the Hamiltonian~\eqref{eq80}  is Hermitian, it should have only real eigenvalues, should it not?''
Our answer is the following: indeed, the Hamiltonian~\eqref{eq80}  is Hermitian in the functional space of the upper half as well as on the real axis of the complex eigen-wave-number plane $K$ (see Figure~\ref{fig2}a), and hence the eigenvalues are real for the states that exist in this Hermitian region;
the Hamiltonian is, however, generally non-Hermitian in the functional space of the lower half of the complex eigen-wave-number plane $K$, and therefore the eigenvalues can be complex for the states there, namely the resonant and anti-resonant states (as well as the anti-bound states)~\cite{Hatano08,Hatano09,Hatano10}.
The existence of the complex eigenvalues for these states is thus not a mystery at all.

Further natural questions may be in order here.
What happens to the probability interpretation if the state diverges as $|x|\to\infty$?
Is the probability conserved if the Hamiltonian is non-Hermitian?
To~answer these questions, we next demonstrate that the spatial divergence balances out the temporal decay of the resonant state.

The key is to define a spatially partial ``norm'' of the diverging resonant state~\eqref{eq130} in the form~\mbox{\cite{Hatano08,Hatano09,Hatano10}}
\begin{align}\label{eq310}
\langle\psi_\R|\psi_\R\rangle_L=&
|\psi_0|^2+\sum_{x=-L\atop x\neq 0}^{+L}|\psi_1|^2 e^{- 2|x| \im K_\R},
\end{align}
%% Authors' correction: We noticed the absolute value marks were missing.
where we used the notation
\begin{align}\label{eq320}
|\psi_\R\rangle
=\sum_{x=-\infty}^{-1}\psi_1e^{-i K_\R x}|x\rangle
+\psi_0|0\rangle
+\sum_{x=+1}^{+\infty}\psi_1e^{i K_\R x}|x\rangle
\end{align}
with $0<\re K_\R<\pi$ and $\im K_\R<0$.
The ``norm''~\eqref{eq310} is a finite positive number for a positive integer $L$.
For large $L$, the most dominant term of Equation~\eqref{eq310} is
\begin{align}\label{eq330}
\langle\psi_\R|\psi_\R\rangle_L
%=&|\psi_0|^2+2|\psi_1|^2e^{- 2 \im K_\R}
%\frac{e^{- 2L \im K_\R}-1}%
%{e^{- 2 \im K_\R}-1}
\simeq |\psi_1|^2
\frac{e^{- 2L \im K_\R}}%
{\sinh (-\im K_\R)}
\end{align}

Remember here that since $\re K_\R>0$ for the resonant state (see Figure~\ref{fig2}a), it has only outgoing waves; that is, the probability is leaking from the range $-L\leq x\leq +L$ and flowing out towards $x\to\pm\infty$.
Correspondingly, the partial ``norm'' of the time-dependent part of the resonant state,
\begin{align}
|\Psi_\R\rangle=|\psi_\R\rangle e^{-it E_\R/\hbar},
\end{align}
decays in time as 
\begin{align}\label{eq340}
\langle\Psi_\R|\Psi_\R\rangle_L=\langle\psi_\R|\psi_\R\rangle_L\ 
e^{2 t\im E_\R/\hbar}
\end{align}
with $\im E_\R<0$ for the resonant state (see Figure~\ref{fig2}b).

We now demonstrate that the probability flow out of the range $-L\leq x\leq +L$ is equal to the temporal decay in Equation~\eqref{eq340}:
\begin{align}\label{eq345}
\frac{d}{dt}\langle\Psi_\R|\Psi_\R\rangle_L=
\frac{2}{\hbar} 
\langle\psi_\R|\psi_\R\rangle_L\ 
\im E_\R\ e^{2 t\im E_\R/\hbar}.
\end{align}
For the probability flow, we compute the expectation value of the local current at $x=\pm L$.
The standard expression of the current in a continuous space in one dimension is given by
\begin{align}
J(x)&=\re \left(\varphi(x)^\ast\frac{p}{m}\varphi(x)\right)
=\frac{\hbar}{2im}\left(\varphi(x)^\ast\frac{d}{dx}\varphi(x)-\varphi(x)\frac{d}{dx}\varphi(x)^\ast\right).
\end{align}
%provided the usual normalization.
In the discretized space of the state~\eqref{eq260}, the local current is therefore defined by
\begin{align}
J(x)=&-\frac{i}{\hbar}
\left[{\varphi_x}^\ast \left(\varphi_{x+1}-\varphi_x\right)-\varphi_{x}\left({\varphi_{x+1}}^\ast-{\varphi_x}^\ast\right)\right]
\\
=&\frac{i}{\hbar}\left({\varphi_{x+1}}^\ast\varphi_{x}-\varphi_{x+1}{\varphi_x}^\ast\right)
\end{align}
for $x>0$, while $J(x)=-J(-x)$ for $x<0$ because of the spatial symmetry.
%with a possible ambiguity of $1/2$ for the spatial index $x$.
%% Authors' correction: We changed the text here.
%again provided the usual normalization.
For the resonant state~\eqref{eq320}, 
%we should explicitly write out the normalization; 
we thereby find the current at the right boundary $x=L$ for large $L$ in the form
\begin{align}\label{eq400}
J_\R(L)=&\frac{i|\psi_1|^2}{\hbar}\left[e^{-i{K_\R}^\ast(L+1)+i{K_\R}L}-e^{-i{K_\R}^\ast L+i{K_\R}(L+1)}\right]
\nonumber\\
\simeq&\frac{2|\psi_1|^2}{\hbar}%\sinh(- \im K_\R)
e^{-2L\im {K_\R}}\sin \re {K_\R}>0
%=&\frac{i|\psi_1|^2}{\hbar\langle\psi_\R|\psi_\R\rangle_L}\left[e^{-i{K_\R}^\ast (L-1)+i{K_\R}L}-e^{-i{K_\R}^\ast L+i{K_\R}(L-1) }\right]
%\nonumber\\
%=&-\frac{2}{\hbar}|\psi_1|^2e^{-(L-1)\im {K_\R}}\sin \re {K_\R}
\end{align}
and the one at the left boundary $x=-L$ in the form $J_\R(-L)= -J_\R(L) <0$. % because of the spatial symmetry,
% where we adjusted the ambiguity of $1/2$ for $L$.
%% Authors' correction: We removed a part of the sentence here.
The sum of these flows is the total probability flow out of the range $-L\leq x\leq +L$.
Taking account of the temporal decay of the all wave amplitudes, we arrive at the estimate
\begin{align}\label{eq410}
\left(J_\R(L)+\left|J_\R(-L)\right|\right)e^{2 t\im E_\R/\hbar}.
\end{align}

On the other hand, the time derivative in Equation~\eqref{eq345} is estimated as follows. We note that
\begin{align}
E_\R=&-2\cos K_\R
\\
=&-2\cos \re K_\R
\cosh\im K_\R
+2i \sin  \re K_\R
\sinh\im K_\R,
\end{align}
and hence $\im E_\R=2\sin  \re K_\R \sinh\im K_\R$.
Multiplying this with Equation~\eqref{eq330}, we evaluate the right-hand side of Equation~\eqref{eq345} as in
\begin{align}\label{eq440}
\frac{2}{\hbar} \langle\psi_\R|\psi_\R\rangle_L\im E_\R\ e^{2 t\im E_\R/\hbar}
\simeq &
-\frac{4|\psi_1|^2}{\hbar} e^{-2L\im {K_\R}}
\sin \re  K_\R\ e^{2 t\im E_\R/\hbar}.
\end{align}
Comparing Equation~\eqref{eq440} with Equation~\eqref{eq400} accompanied by Equation~\eqref{eq410}, we arrive at the relation
\begin{align}\label{eq450}
-\frac{d}{dt}\langle\Psi_\R|\Psi_\R\rangle_L
=\left(J_\R(L)+\left|J_\R(-L)\right|\right)e^{2 t\im E_\R/\hbar},
\end{align}
which demonstrates that the temporal decrease of the ``norm'' over the region $-L\leq x\leq +L$ flows out of the region~\cite{Hatano08,Hatano09,Hatano10}.

As the answer to the question ``Is the probability conserved if the Hamiltonian is non-Hermitian?'',
the probability of the resonant state is conserved in a particular way in which the temporal exponential decay is negatively equal to the spatial exponential divergence~\cite{Hatano08,Hatano09,Hatano10}.
For the question ``What happens to the probability interpretation if the state diverges as $|x|\to\infty$?'',
we emphasize that the spatial divergence of the resonant state is not ``unphysical'' at all;
on the contrary, it is \textit{necessary} for the probability conservation~\cite{Hatano08,Hatano09,Hatano10}.

This concludes our discussion on the solutions of the time-independent Schr\"{o}dinger Equation~\eqref{eq50} with a time-reversal symmetric and seemingly Hermitian Hamiltonian.
To summarize the discussion, the equation can have eigenstates that break the time-reversal symmetry, namely the resonant states that decay in time and the anti-resonant states that grow in time, both of which we showed are very much physical.
In the next section, we finally show how these symmetry-breaking states emerge in the dynamics of the time-dependent Schr\"{o}dinger Equation~\eqref{eq10}, namely in the phenomenon called the arrow of time.

\section{One More Element for ``Arrow of Time''}
\label{sec5}

So far we have shown that the Hamiltonian of an open quantum system can have eigenstates that break time-reversal symmetry. This is a necessary condition for the existence of an ``arrow of time''. However, we will argue that it is not a sufficient condition. Indeed, the arrow of time means that we observe specific symmetry-breaking states that evolve in the same direction ({e.g.}, resonant states decaying as time increases), while we do not observe states that evolve in the opposite direction ({e.g.}, anti-resonant states growing as time increases).  

This point is lacking in most of the arguments about the arrow of time. 
For example, there is an argument on the Poincar\'{e} recurrence time~\cite{Chandrasekhar43}, which is the time after which a closed system comes back to the initial state; it has been found to become exponentially large as the system size increases~\cite{Venuti15,Zhang17,Anishchenko18}, and hence only the initial part of the evolution is observable. 
This actually may correspond to the fact that the system of infinite size has decaying resonant states and growing anti-resonant states as eigenstates;
in a large but finite system, the outgoing flow generated by a~decaying state would be bounced back and the initial state would recur but only after a long time, which is indeed the physics of the Poincar\'{e} recurrence time.
Therefore, the divergence of the recurrence time in the infinite system indeed separates the decaying states from the growing states.

However, the argument of the Poincar\'{e} recurrence time does not logically say that a decaying state should be chosen for the initial part of the time evolution; according to the argument, the choice of a growing anti-resonant state would be equally plausible.
Why are the anti-resonant states, which should grow in time, not chosen in  positive times and why do the resonant states, which should decay in time, dominate the dynamics in the direction of future?
The argument of the recurrence time falls short of a satisfactory explanation of the arrow of time in this respect.

We can see a similar problem in the introduction of an infinitesimal constant in computing the retarded and advanced Green's functions. 
One adds an infinitesimal constant to shift the eigenvalues on the real axis to the lower half of the complex energy plane in order to find the retarded Green's function, which has only outgoing waves from the source.
One also shifts the eigenvalues to the upper half in order to find the advanced Green's function, which has only incoming waves to the sink.
One~then typically legitimatizes the shifts of the eigenvalues only on the basis of the fact that the resulting Green's function seems correct;
it again falls short of a satisfactory explanation of the arrow of time.
In the eigenvalue problem of the Hamiltonian~\eqref{eq80}, the infinitesimal shift to the lower half of the complex energy plane is substituted by the explicitly negative imaginary part of the resonant eigenvalue and that to the upper half by the explicitly positive imaginary part of the anti-resonant eigenvalue.
They forms a time-reversal symmetric pair, and hence any time-reversal symmetric argument does not give a reason to choose one pair over the other. 
%The present argument shows the reason why the eigenvalues in the lower half should give the decay and those in the upper half should give the growth.
%As is indicated in Fig,~\ref{fig2}(b), the eigenvalues in the lower half of the complex energy plane are the resonant states, for which the temporal decay is equal to the probability flow out of the central region, as was demonstrated by Equation~\eqref{eq450}.
%We can also show that the temporal growth of anti-resonant states is equal to the probability flow into the central region as the complete time reversal of Equation~\eqref{eq450}.
%We will show below that the former indeed excels in the direction of future.

To have an arrow of time oriented towards the future, we need a mechanism
that suppresses anti-resonant states, making the resonant states dominant for positive times.  We will show that this situation arises naturally when we consider an initial-condition problem: given an initial state at $t=0$, the problem is to calculate the evolving state for $t>0$. We will show that, for the present model, we can decompose the time-evolving state into two components, one associated with the resonant state and the other one associated with the anti-resonant state. If the initial state is time-reversal symmetric, its resonant and anti-resonant components have equal weights at $t=0$. However, for $t>0$ the time evolution operator  naturally suppresses the anti-resonant states, thereby producing an arrow of time in our simple model.

It is important to remark that an arrow of time oriented towards the past is also possible, to comply with the time-reversal symmetry of the Schr\"{o}dinger equation~\eqref{eq10}. This may be difficult for us to imagine, but a past-oriented arrow of time can in fact be observed as follows: we must consider a~final-condition problem, where the time-reversal invariant state at $t=0$ is considered a~final condition, and we select time-dependent states for $t<0$ that lead to the final condition. In this case, the anti-resonant components of the selected states are dominant, while the resonant ones are suppressed. 
Indeed, this procedure was implemented in the post-measurement data analysis of an experiment of the weak measurement~\cite{Foroozani16}.

The nontrivial part of our argument is the decomposition of states into resonant and anti-resonant states that are naturally suppressed for $t<0$ and $t>0$, respectively. This decomposition is not a~simple resolution of the time-evolution operator in terms of all the eigenstates of the Hamiltonian (including states with discrete and continuous spectrum). Instead, it is a resolution that involves only the eigenstates of the Hamiltonian with discrete spectra~\cite{Hatano14,Ordonez17a,Ordonez17b}. This decomposition is explicitly time-reversal invariant, because it includes complex-conjugate pairs of resonant and anti-resonant~states. 

As an example, we consider the survival amplitude $A(t) = \langle 0| e^{-i H t/\hbar} |0\rangle$ of the time-reversal-\ symmetric initial state $|0\rangle$. 
We can compute $A(t)$ itself in a standard way;
we here show how the amplitude is broken down into resonant and anti-resonant components.
The time-reversal invariant decomposition takes the form
%%%
\begin{align} \label{eq:xrep}
 A(t) &=\sum_{n}\langle 0|\chi_n(t)\rangle
\end{align}
%%%
where the $n$ subindex denotes either ``resonant'' or ``anti-resonant''. The resonant and anti-resonant components are given by~\cite{Hatano14,Ordonez17a,Ordonez17b}
%%%
\begin{align} \label{eq:Xin}
 |\chi_n(t) \rangle &=\frac{1}{2\pi i}
\int_{C}d\lam\,\left(-\lam+\frac{1}{\lam}\right)\exp\left[i\left(\lam+\frac{1}{\lam}\right)t\right]
%\nonumber\\
%&\times
|\psi_n\rangle\frac{\lam_n}{\lam-\lam_n}\langle\tilde{\psi}_n|0\rangle.
\end{align}
%%%
where $C$ denotes a clockwise contour of the unit circle in the complex $\lambda$ plane, $\lambda_n$ are given by Equation~(\ref{eq:lambdap}), and $\langle\tilde{\psi}_n|$ are the left eigenstates corresponding to the right eigenstates $|\psi_n\rangle$. 

\begin{figure}
\centering
\includegraphics[width=0.55\textwidth]{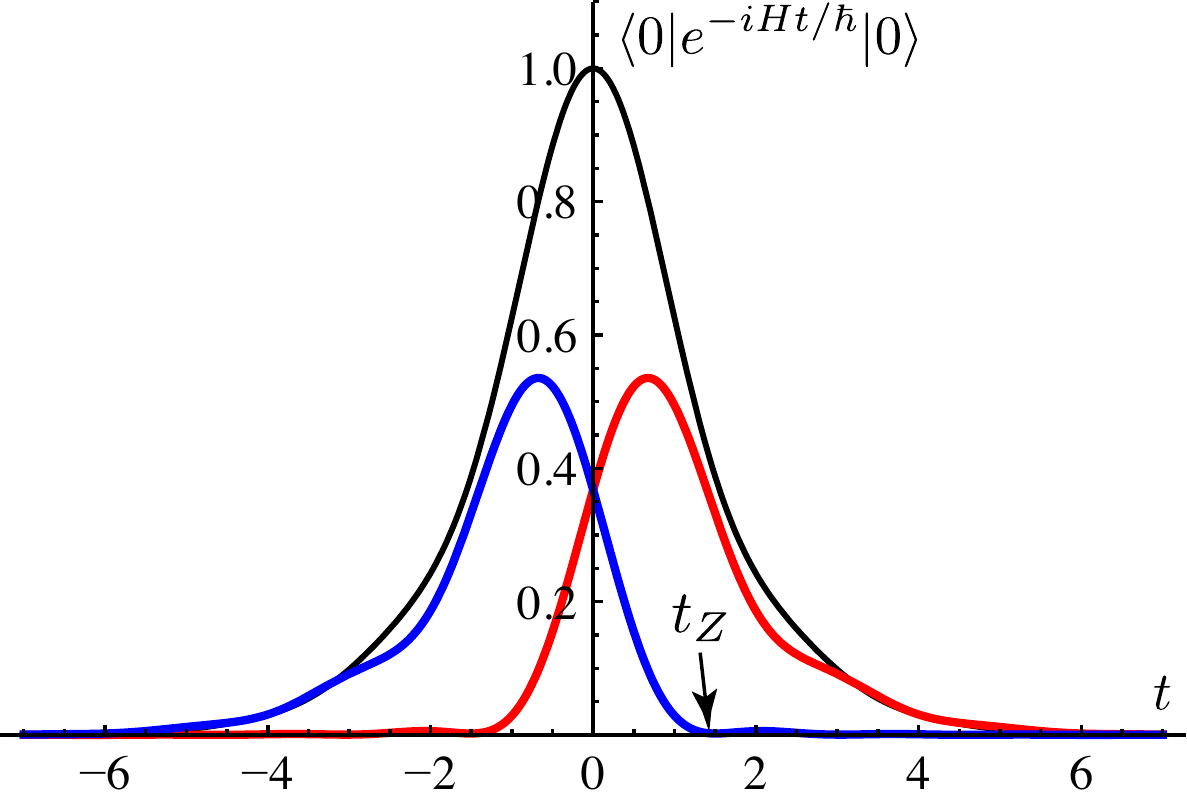}
\caption{Total survival probability $|A(t)|^2$  (black symmetric line with the highest peak); the resonant state $|\langle 0|\chi_\R(t)\rangle|^2$ component (red line with a peak occurring at a positive time ); the anti-resonant state $|\langle 0 |\chi_{\AR}(t)\rangle|^2$ component (blue line with peak occurring at a negative time). The parameters we used are $t_0  = 0.5$ and $V_0=-0.8$.}
\label{fig3}
\end{figure}

Figure~\ref{fig3}~\cite{Ordonez17b} shows the numerically evaluated components $|\langle 0|\chi_n(t)\rangle|^2$ of the survival probability corresponding to the resonant and anti-resonant states. At $t=0$ both components have equal weights. (Since we plot the square modulus of each component and the components are complex numbers, the total probability at $t=0$ is not twice as large as the probability of each individual component). The figure clearly shows that the anti-resonant component is suppressed for $t>0$.
(In fact, it is suppressed after a short  time $t_Z$, which we have associated with the ``Zeno time''  in Ref.~\cite{Ordonez17a}).  As the anti-resonance component is suppressed, the resonant component  alone approaches the complete survival probability, which decays exponentially by then. Conversely, the resonant component is suppressed for $t<0$.
%%%
%%%

\section{Summary and Discussion}
\label{sec6}

We have argued that  the arrow of time in open quantum systems emerges naturally in terms of resonant and anti-resonant state components of quantum states. The following are the main points of our argument:
\begin{enumerate}[leftmargin=*,labelsep=4.9mm]
\item The Hamiltonian of an open quantum system can have complex eigenvalues corresponding to unnormalizable eigenstates. These are the resonant and anti-resonant states, which decay towards the future and towards the past, respectively. Since we assume that the Hamiltonian is time-reversal symmetric, these eigenstates always appear in complex-conjugate pairs.
\item Even though the resonant and anti-resonant states are unnormalizable, there is a physical interpretation of their exponentially growing wave functions: their exponential growth in space counters their exponential decay in time, to ensure probability conservation.
\item An evolving state can be decomposed into into resonant-state components and antiresonant-state components, using an explicitly time-reversal symmetric decomposition of the time-evolution operator, Equation \eqref{eq:Xin}. (In general, there may also be bound-state and antibound-state components). In our simple model, there is only one resonant state and one anti-resonant~state. 
\item For an initially time-reversal symmetric state, the resonant and anti-resonant components have equal weights. However, for positive times, the time evolution operator suppresses the anti-resonant component, so that the resonant component becomes dominant. This leads to the emergence of a future-oriented arrow of time, because the resonant state is not time-reversal invariant by itself. Similarly, a past-oriented arrow of time emerges for negative $t$.
\end{enumerate}

Our approach has some similarity with the work of Prigogine and collaborators \cite{Prigogine73, Prigogine81, Bohm89a, Petrosky97, Driebe1999}. In their approach, the arrow of time is directly connected to the sign of the imaginary part of the poles of the resolvent (or the Green's function). For example, a future-oriented arrow of time corresponds to poles with the negative imaginary parts, {i.e.}, with the resonances. The difference between their approach and ours is that they identify the irreversible dynamics with the subspace spanned by the resonant eigenstates, and hence introduce an arrow of time \textit{a priori}. Meanwhile, we use an explicitly time-reversal symmetric expansion, which includes both resonant and anti-resonant states. This expansion shows that, starting with a time-reversal symmetric state, the time evolution breaks this symmetry and leads to the emergence of the arrow of time as explained in the point $4$ above.

We remark that the exponential decay of our model is only one of the irreversible phenomena displaying an arrow of time. Other irreversible phenomena include approaches to thermal equilibrium or decoherence \cite{Zurek}. The irreversibility associated with both is, in fact, connected to resonances \cite{Petrosky97, Barsegov}. Therefore, we expect that there is a breaking of resonance-antiresonance symmetry behind these phenomena as well.

\begin{figure}[H]
\centering
\includegraphics[width=0.45\textwidth]{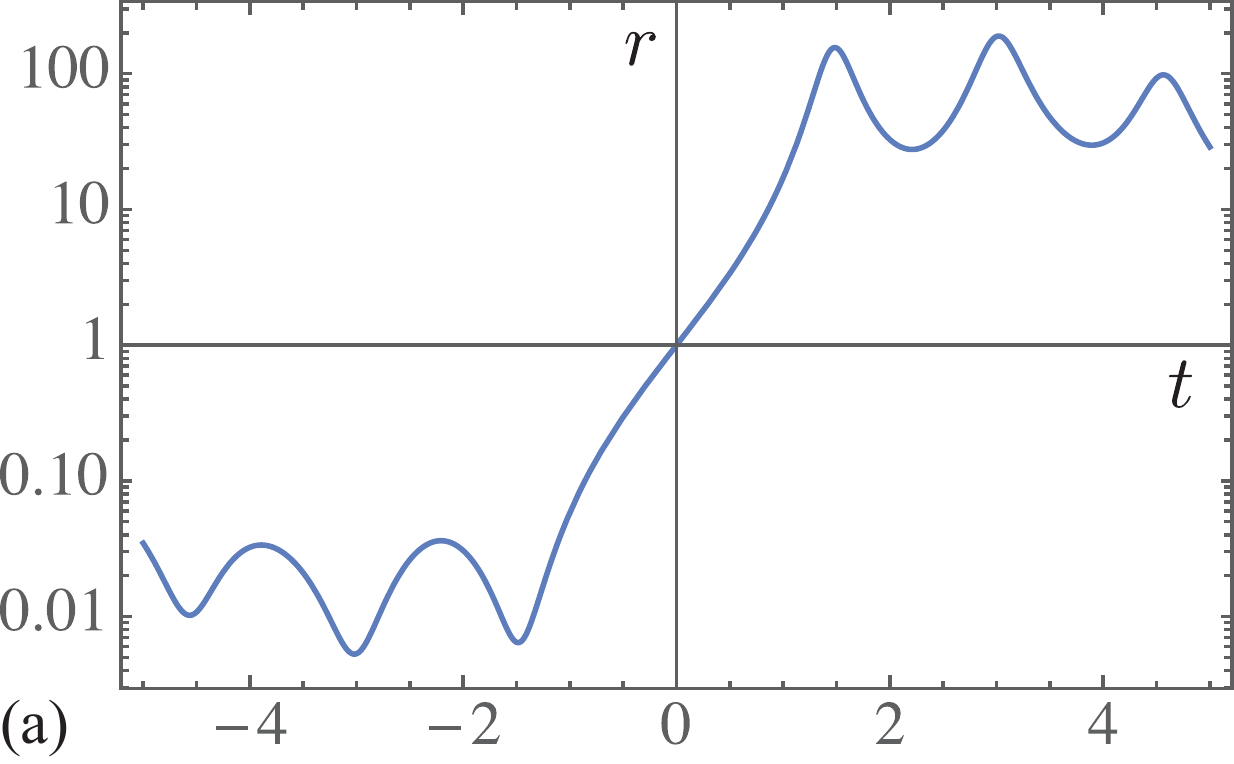}
\hspace{0.05\textwidth}
\includegraphics[width=0.45\textwidth]{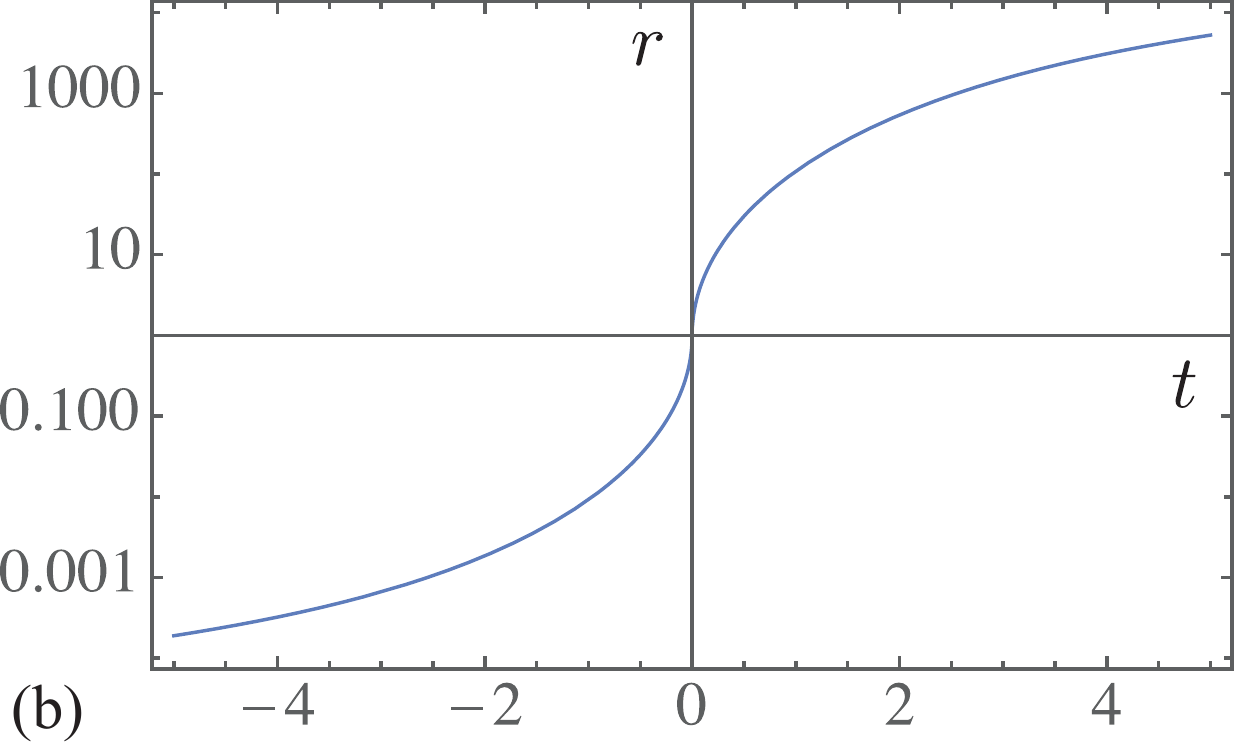}
\vspace{0.5\baselineskip}
\caption{Ratio of the resonant to anti-resonant components of ({\bf a}) the survival probability plotted in Figure~\ref{fig3} and ({\bf b}) the survival probability for the continuous model of Ref.~\cite{Ordonez17a}.}
\label{fig4}
\end{figure}

Breaking of the symmetry between resonant and anti-resonant components of the initial state depends on the sign of $t$. Positive $t$ favors the resonant component, while negative $t$ favors the anti-resonant component.  This is somewhat analogous to the magnetization of a ferromagnetic material by an external magnetic field. The direction of the external field is analogous to the sign of $t$, whereas the magnetization is analogous to relative weight of the resonant and anti-resonant components of the evolving state. 
We may quantify this relative weight by the ratio
\begin{align}
r(t)=\frac{|\langle 0 |\chi_{\R}(t)\rangle|^2}{|\langle 0 |\chi_{\AR}(t)\rangle|^2}
\end{align}
of these components~\cite{Ordonez17a},  which we plot in Figure~\ref{fig4}a. Since the present model is discrete in space, the ratio $r(t)$ contains no singularities or critical exponents in this case. However, if we plot the same ratio for the continuous model considered in Ref.~\cite{Ordonez17a}, we obtain the graph in Figure~\ref{fig4}b, which contains an infinite slope at $t=0$; more specifically, it varies as $|t|^{-1/2}$ as $t$ approaches $0$. For future work, it is worth exploring this behavior further, as well as other possible analogies between phase transitions and the appearance of the arrow of time in open quantum systems.
%%%
\authorcontributions{The present two authors equally contributed to the foundation of the theory reviewed here.}  %please add it.
%% Authors' Reply: Here is our contribution breakdown.

%%%%%%%%%%%%%%%%%%%%%%%%%%%%%%%%%%%%%%%%%%
\funding{N.H.'s work was partially supported by JSPS Grant-in-Aid for Scientific Research (B) Grant Nos.~22340110, 15K05200, 15K05207, and 26400409 from Japan Society for the Promotion of Science, a Research Grant from the Yamada Science Foundation, a Research Grant in the Natural Sciences from the Mitsubishi foundation, as well as the Holcomb Awards Committee and Woods Lecture Series at Butler University and the Clark Way Harrison Visiting Professorship of Washington University in St.~Louis. G.O.~acknowledges the Institute of Industrial Science at the University of Tokyo, the Department of Physical Science at Osaka Prefecture University, the Holcomb Awards Committee and the LAS Dean's office at Butler University for support of this work.} %Please disclose any funding information, or add "This research received no external funding."
%% Authors' Reply: Thank you for the reminder.

%%%%%%%%%%%%%%%%%%%%%%%%%%%%%%%%%%%%%%%%%%
\conflictsofinterest{The authors declare no conflict of interest.}
%Please disclose conflicts of interest, or add “The authors declare no conflicts of interest.”
%% Authors' Reply: Thank you for the reminder.

%%%%%%%%%%%%%%%%%%%%%%%%%%%%%%%%%%%%%%%%%%
\acknowledgments{We thank Savannah Garmon, Tomio Petrosky, and Satoshi Tanaka for insightful discussions and helpful suggestions throughout the work.}
\reftitle{References}

%%%%%%%%%%%%%%%%%%%%%%%%%%%%%%%%%%%%%%%%%%
\end{document}